\documentclass[aps,graphicx,twocolumn,epsfig,showpacs]{revtex4}

\usepackage{epsfig}

\begin{document}

\author{C. Menotti}
\author{S. Stringari}
\affiliation{CNR-INFM BEC and Dipartimento di Fisica, Universit\`a
di Trento, I-38050 Povo, Italy}

\title{Detection of Pair-Superfluidity for bosonic mixtures in optical lattices}

\begin{abstract}
We consider a mixture of two bosonic species with tunable
interspecies interaction in a periodic potential and discuss the
advantages of low filling factors on the detection of the
pair-superfluid phase. We show how the emergence of such a phase
can be put dramatically into evidence by looking at the
interference pictures and density correlations after expansion and
by changing the interspecies interaction from attractive to
repulsive.
\end{abstract}


\maketitle

Ultra-cold atoms in optical lattices are presently one of the best
environments for the study of exotic quantum phases
\cite{lewenstein2007, bloch2008}. The experimental demonstration
of the superfluid to Mott transition  \cite{greiner2002} has
opened the way to the study of strongly correlated phases in
lattices. In the case of mixtures of different atomic species or
different internal levels, new phenomena related to quantum
magnetisms and spin physics arise \cite{lewenstein2008}.

The quantum phase that is central to this work is the
pair-superfluid (PSF) phase for a mixture of two different bosonic
species. This phase consists in the formation of a superfluid of
pairs where atoms of different species preferentially hop together
in the lattice. It is characterised by the vanishing of the
single-species order parameter and by the emergence of a
pair-order parameter. This problem is related to the historical
question formulated, e.g., by Nozi\`eres and Saint James
\cite{nozieres1982}, concerning the impossibility of coexistence
of single and pair superfluidity. This impossibility lies in the
fact that a single species condensate is associated with the
macroscopic occupation of the zero momentum (quasi-momentum) state
for each species, while a pair superfluid requires the macroscopic
occupation of the zero relative momentum, without any constraint
on the value that the single species momenta can separately take.
As a consequence, single species condensation is destroyed. In
this work, we will see how this picture is directly reflected into
physical quantities accessible in experiments.

The existence of a pair-superfluid phase in lattices has been
already pointed out in several papers \cite{kuklov2003,
altman2003, kuklov2004, arguelles2007, soyler2008, hubener2009,
mathey2009, hu2009}. The main emphasis has been put on the
situation of equal density for the two species leading to total
integer and half integer filling factor.  For integer filling
factors, PSF arises in the regime where the interspecies
interaction almost completely compensates the repulsive
intraspecies interaction. Unfortunately,
 the precision on the values of the interaction
strengths and the very small values of the tunneling parameter
required to get pair superfluidity, makes this phase almost
unaccessible experimentally. At half filling factor for each
species (total integer filling), the PSF phase (analogous to the
$x-y$ ferromagnet) is predicted in a larger region of the phase
space. At low tunneling and in the presence of asymmetries between
the interaction and tunneling parameters of the two species, the
PSF phase competes with the insulating-like anti-ferromagnetic
ordering. Instead for total incommensurate filling factor, no
insulating-like phases (Mott or antiferromagnetic-like) exist. In
this regime, easily accessible signatures for the experimental
observation of pair superfluidity are available. In this paper, we
would like to complement the predictions in \cite{hu2009},
discussing the role played by the two-body momentum distribution
and commenting on the effect of interactions in the expansion.

At total incommensurate filling factor and zero temperature, two
important phases are naturally conceived \cite{footnote}: $(i)$ a
double superfluid (2SF), where both species are independently
superfluid and single-species coherence exists, and $(ii)$ a PSF
phase, characterized by pair coherence. Assuming that interactions
do not affect significantly the expansion of the atomic cloud
after release, all required information needed to distinguish
between the two phases are included in the pictures of the two
species after expansion: first of all, the interference fringes,
typical of single-species coherence, will appear for the 2SF phase
and vanish in the case of PSF. Moreover, the density-density
correlations between the two species after expansion carry
information about the correlations in momentum space before
expansion, which are dramatically different for the two phases.

We consider two bosonic atomic species in a lattice, described by
the Bose-Hubbard Hamiltonian

\begin{eqnarray}
\label{GBH} H&=&- \sum_{\langle ij \rangle}
\left[J_a a_i^\dag a_j + J_b b_i^\dag b_j \right]+ \\
&&+\sum_{i,\sigma} \left[\frac{U_\sigma}{2} n_i^\sigma (n_i^\sigma
-1) \right] + \sum_i U_{ab} n_i^a n_i^b, \nonumber
\end{eqnarray}
where $\sigma=a,b$ indicates the two bosonic species,
$a_i,b_i,a_i^\dag,b_i^\dag$, and $n^\sigma_i$ are respectively the
annihilation, creation operators and the density of species $a$
and $b$ at site $i$. The notation $\langle i j \rangle$ represents
nearest neighbors. The intraspecies on-site interactions
$U_\sigma$ and tunneling parameters $J_\sigma$ depends in the
standard way on the optical lattice potential and scattering
lengths. Generally speaking the most favorable conditions for PSF
are given by a complete symmetry between the two species, as far
as interaction, hopping and density are concerned. This is the
situation that we will assume in this paper ($N_a=N_b=N$, $J_a=
J_b=J$, $U_a=U_b=U$), focusing on the possibility of changing the
interspecies interaction $U_{ab}$ from negative to positive by
tuning the interspecies scattering length via a Feshbach resonance
over a wide range, as demonstrated in \cite{catani2008}.

The solution of Hamiltonian (\ref{GBH}) is a non-trivial many-body
task. Quantum Monte Carlo (QMC) calculations provide the most
accurate results \cite{kuklov2003, kuklov2004, soyler2008}.
Evidences of the PSF phase have been recently obtained also by
matrix-product-state \cite{arguelles2007, mathey2009, hu2009} and
dynamical mean-field approaches \cite{hubener2009}. Instead the
standard mean-field approximation based on a generalized
Gutzwiller Ansatz $|\Phi\rangle= \prod_i \sum_{n,m} f_{n,m}^{(i)}
|n_a,n_b\rangle_i$, which in principle includes correlations
between the two atomic species, misses the pair superfluid phase
due to the impossibility of correctly accounting for second order
hopping which is at the origin of the PSF phase. Recently, a
treatment based on a mean-field analysis of the effective
Hamiltonian in the pair-subspace applied to a bilayer system of 2D
dipolar lattice bosons has proven successful to describe the PSF
and pair-supersolid (PSS) phases \cite{trefzger2009}.

In order to capture the basic physics of the pair correlations
underlying the emergence of the PSF, in this work we employ a
toy-model based on the exact diagonalisation of (\ref{GBH}) for a
system of few atoms occupying few lattice wells in 1D with
periodic boundary conditions. Of course sharp phase transitions
are not accounted for by our treatment, but we believe that the
main conclusions remain valid also for larger systems and higher
dimensions.

The comparison between filling factor equal to and less than $1/2$
is useful, since  for filling factor exactly equal to $1/2$, there
exist a particle-hole symmetry between positive and negative
interspecies interaction in the almost hard-core limit
$|U_{ab}|\ll U$, leading to the pairsuperfluid phase (PSF) for
attractive interactions (pairing of two atoms of different
species) and the so-called countersuperfluid phase (SCF) for
repulsive interactions (pairing of one atom with a hole of the
other species). Instead, for equal filling factor less than 1/2
for each species, PSF still persist, but SCF pairing is
suppressed. This different behaviour for positive and negative
interspecies interaction might help identifying the formation of
PSF as discussed below.

In order to put into evidence the differences between filling
factor $\nu=1/2$ and $\nu<1/2$, we will consider a system of 4
wells and 2 atoms of each species ($N_w=4$, $N=2$) and a system of
6 wells and 2 atoms of each species ($N_w=6$, $N=2$), which are
the smallest cases including the possibility of having equal
filling factors $\nu \le 1/2$ and non trivial on-site
interactions. We look at the following quantities
\cite{footnote2}:

\begin{eqnarray}
&&{\cal V}_{2SF}=\langle a^{\dag}_i a_{i+1} \rangle = \langle
b^{\dag}_i b_{i+1} \rangle, \label{V_2SF} \\
&&{\cal V}_{PSF}=\langle a^{\dag}_i b^{\dag}_i a_{i+1} b_{i+1}
\rangle-
\langle a^{\dag}_i a_{i+1} \rangle \langle b^{\dag}_i b_{i+1} \rangle, \label{V_PSF} \\
&&{\cal V}_{SCF}=\langle a^{\dag}_i b_i a_{i+1} b^{\dag}_{i+1}
\rangle- \langle a^{\dag}_i a_{i+1} \rangle \langle b_i
b^{\dag}_{i+1} \rangle. \label{V_SCF}
\end{eqnarray}
The quantities ${\cal V}_{2SF}$, ${\cal V}_{PSF}$ and ${\cal
V}_{CSF}$ characterise respectively the 2SF, PSF and SCF phases.
We also consider single particle and two-particle momentum
distributions in order to make a useful link to the experiments.

\begin{center}
\begin{figure}[t]
\includegraphics[width=0.85\linewidth]{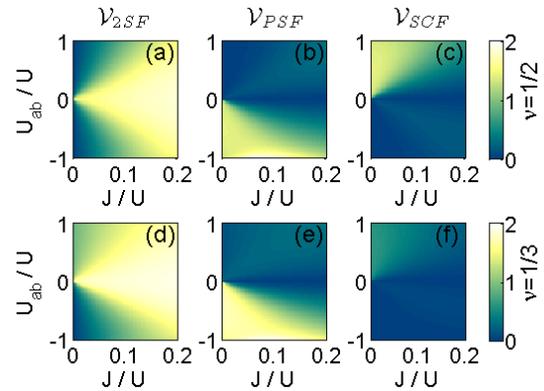}
\caption{(Color online) Phase diagram for $N_w=4,6$ (upper and
lower row, respectively) and $N=2$. (a,d) Single-particle
coherence $N_w {\cal V}_{2SF}$, as defined in Eq.(\ref{V_2SF});
(b,e) Pair coherence $N_w {\cal V}_{PSF}$, as defined in
Eq.(\ref{V_PSF}); (c,f) Counter-pair coherence $N_w {\cal
V}_{SCF}$, as defined in Eq.(\ref{V_SCF}). The factor $N_w$ allows
a better comparison between the two different lattice sizes. The
region of parameters of strong attractive interaction $U_{ab}<-U$
corresponds to collapse in a large system.} \label{fig1}
\end{figure}
\end{center}

In Fig.\ref{fig1}, we show the phase diagram  as a function of $J$
and $U_{ab}$. One can clearly identify the PSF (or SCF) regions as
the dark regions in Fig.\ref{fig1}(a,d) where single-species
coherence vanishes, and at the same time pair-coherence (or
counter-pair coherence) is different from zero, namely the light
regions in Fig.\ref{fig1}(b,e) (or Fig.\ref{fig1}(c) for SCF). As
explained above, for $\nu=1/2$, PSF and SCF are found respectively
for attractive and repulsive interaction. Instead, for $\nu<1/2$,
SCF is absent.
For low filling, the crossover from 2SF to PSF is governed by the
competition between the single particle hopping $J$ and the energy
cost for breaking a pair, equal to $|U_{ab}|$. For that reason,
PSF is found for attractive interspecies interactions at
sufficiently low tunneling parameter $J \ll |U_{ab}|$.

\vspace{-0.9cm}

\begin{center}
\begin{figure}[t]
\includegraphics[width=0.85\linewidth]{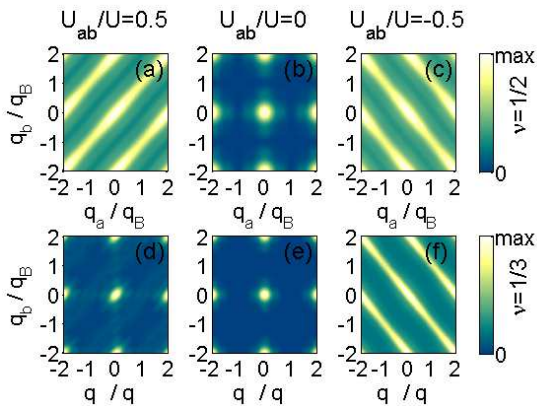} \caption{(Color
online) Two-body momentum distribution $n^{(a,b)}(q_a,q_b)$ for
$N=2$ and $N_w=4,6$, i.e. filling factor $1/2$ and $1/3$ (upper
and lower row, respectively). (a,d) repulsive interparticle
interaction $U_{ab}=0.5U$; (b,e) vanishing interparticle
interaction $U_{ab}=0$; (c,f) attractive interparticle interaction
$U_{ab}=-0.5U$. In all pictures $J=0.01U$. Two Brillouin zones are
shown for clarity.} \label{fig2}
\end{figure}
\end{center}

An important quantity accessible in experiments, which would
provide an unquestionable proof of PSF, is the measure of
correlations in the momentum distribution, reflecting the fact
that in the PSF and SCF phases, two atoms of different species
form a pair and condense in the state of total quasi-momentum $q_a
\pm q_b=0$ (respectively for atom-atom and atom-hole pairs), as
shown in Fig.\ref{fig2}(a,c,f). In the case of two non interacting
superfluids ($U_{ab}=0$), the two species have completely
uncorrelated momentum distributions, i.e.
$n^{(a,b)}(q_a,q_b)=n^{(a)}(q_a)\times n^{(b)}(q_b)$, where
$n^{(a)}(q_a)$ and $n^{(b)}(q_b)$ separately present interference
peaks at even multiples of the Bragg momentum $q_B$ (see
Fig.\ref{fig2}(b,e)), as happens for standard single component
condensates \cite{pedri2001, greiner2002}. In the presence of
interspecies interactions $U_{ab} \neq 0$, correlations build up
in a very different way depending on the filling factor and on
whether the interactions are repulsive or attractive. For filling
factors exactly equal to $1/2$, the situation is almost symmetric
for positive and negative $U_{ab}$ upon particle-hole duality for
the different species. The momentum correlations are opposite in
the two cases, as shown in Fig.\ref{fig2}(a,c), showing SCF and
PSF respectively. The 2SF phase is recovered in the vicinity of
vanishing interspecies interactions (Fig.\ref{fig2}(b)). For equal
filling factor smaller than $1/2$, a 2SF is obtained both for
vanishing and repulsive interactions (Fig.\ref{fig2}(d,e)), since
the filling factor of one species does not match the filling
factor of the holes of the other. Hence, only negligible momentum
correlations exist for $U_{ab} \ge 0$. On the contrary, attractive
interactions lead to PSF and very strongly correlate the two
different species ($U_{ab}/U=-0.5$, Fig.\ref{fig2}(f)).

\vspace{-0.9cm}

\begin{center}
\begin{figure}[t]
\includegraphics[width=0.8\linewidth]{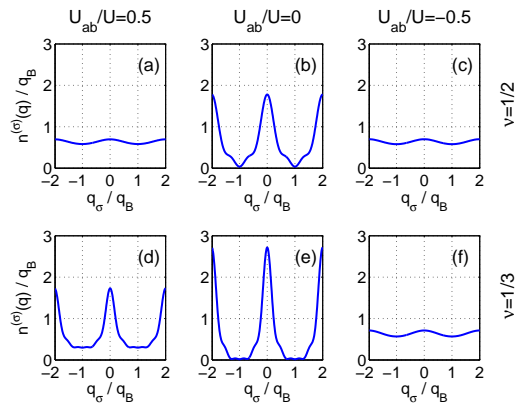} \caption{Single
species momentum distribution
$n^{(\sigma)}(q_\sigma)$ for $N=2$ and $N_w=4,6$, i.e. filling
factor $1/2$ and $1/3$ (upper and lower row, respectively). (a,d)
repulsive interparticle interaction $U_{ab}=0.5U$; (b,e) vanishing
interparticle interaction $U_{ab}=0$; (c,f) attractive
interparticle interaction $U_{ab}=-0.5U$. In all pictures
$J=0.01U$. Two Brillouin zones are shown for clarity.}
\label{fig3}
\end{figure}
\end{center}

The correlations in the two-body momentum distribution are
strictly related to single-particle coherence and strongly affect
the visibility of the single particle momentum distribution, as
shown in Fig. \ref{fig3}. The presence of momentum correlations in
$n^{(a,b)}(q_a,q_b)$ lead to a reduced contrast in
$n^{(\sigma)}(q_\sigma)$. Hence, some signatures of the formations
of the PSF/SCF phases are provided already by the interference in
the single-particle expansion pictures, which is the most easily
accessible experimental method  \cite{hu2009}. For instance, at
low enough tunneling, for $\nu<1/2$ interference is expected at
$U_{ab}\ge 0$, while it disappears (under exactly the same
conditions) by tuning $U_{ab}$ to negative values.

Most important, the single species expansion pictures carry
information about the momentum correlations. In fact, as
demonstrated in \cite{greiner2005}, the direct measurement of the
momentum correlations can be performed by looking at the noise in
the single-species expansion pictures \cite{altman2004,
kuklov2007}. Assuming first that interactions do not affect the
expansion, the single-species densities after time of flight are
given by $n^\sigma_{\rm TOF}(r_\sigma=q_\sigma t
/m_\sigma)=n^{(\sigma)}(q_\sigma)$.  Hence, in the case of PSF,
where the correlations are of the type $q_a+q_b=0$, we expect the
two expansion pictures to be correlated at correctly rescaled
opposite positions and possibly at corresponding points in
different Brillouin zones.

\vspace{-0.9cm}

\begin{center}
\begin{figure}[t]
\includegraphics[width=0.85\linewidth]{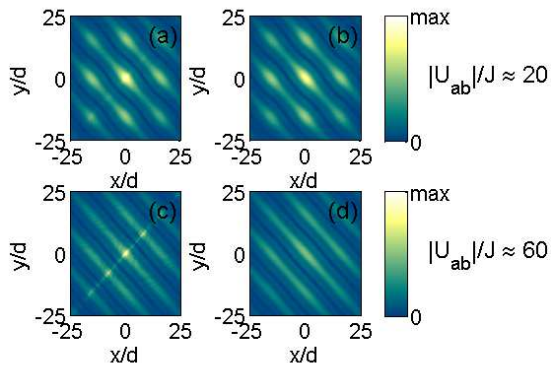} \caption{(Color online) Effect of
attractive interspecies interaction $U_{ab}$ on the expansion.
(a,c) Two-body density after expansion after a time of flight $E_r
t_{\rm TOF} \approx 12$, where $E_r$ is the recoil energy, in the
presence of interspecies interaction $U_{ab}$; (b,d) Free
expansion at the same time of flight (shown for comparison). The
strengths of interaction are $|U_{ab}|/J\approx 20$ in (a,b) and
$|U_{ab}|/J\approx 60$ in (c,d) respectively.} \label{fig4}
\end{figure}
\end{center}

 In usual experiments, interatomic interactions are
not turned off during the expansion, and their effect is
considered to be negligible due to the higher energy scales
involved in the problem. In the present case, it would be a safe
procedure to tune at least $U_{ab}$ to zero just before releasing
the cloud. However, this might not be so easily achievable in an
experiment. For this reason, we have performed simple numerical
estimations of the effect of the interspecies interactions on the
expansion for two atoms released from a 4-well lattice (see
Fig.\ref{fig4}). While the effect of relatively weak attractive
interactions ($|U_{ab}|/J\approx 20$, see Fig.\ref{fig4}(a,b)),
the two-body momentum distribution is hardly modified during the
expansion, we have seen that the two-body momentum distribution
can be affected by attractive interspecies interaction $U_{ab}$,
for interaction strengths leading to pairing ($|U_{ab}|/J\approx
60$, see Fig.\ref{fig4}(c,d)). This effect tends to create also
some correlations along the diagonal $q_a=q_b$, but does not
destroy the correlations typical of the PSF phase (at $q_a=-q_b$).
Hence, recovering the two-body density after expansion via noise
correlation measurements, the distinction between PSF and 2SF
phases still remains very clear. More dramatic effects of the
interactions during the expansion take place only for values of
$|U_{ab}|/J$ which are much beyond the estimated onset of the PSF
phase transition.

\vspace{-0.9cm}

Small asymmetries in the Hamiltonian parameters of the two species
do not destroy the PSF phase. Instead, an unbalance in the
densities of atoms $a$ and $b$ hinders the formation of the pairs.
Unfortunately, we are not able to quantify the realistic effect of
the unbalance due to the small number of atoms considered.
However, one can think of experimental procedures to create a
sample with almost exactly equal populations of the two species.
For instance, one could start with a Mott insulator at unit
filling for both species, and then tailor the optical potentials,
introducing a second laser at half the wavelength, such to split
each lattice well into two equal ones. Applying this procedure
along two dimensions would lead to a filling factor of exactly
$1/4$ for each species. This procedure should remain valid also in
the presence of an external harmonic confinement, especially in
the case of very low tunneling, where the central Mott region at
unitary filling is dominant with respect to the outer superfluid
shell. Alternatively, one can create a unitary filled Mott region
for both species in the trap center and then release the harmonic
trap till the desired filling factor is reached. This method would
favor the creation of the pairs in the Mott phase, which can then
become superfluid once the filling factor is made incommensurate.

The physical ingredient on which pair superfluidity relies is the
second order hopping of two atoms of the different species at
once. This is closely related to the exchange interaction, whose
observability has been recently demonstrated in
\cite{trotzky2008}. The fact that pair hopping is a second order
process in $J$, where $J$ is assumed to be small, might seem to be
discouraging for the experimental observation of the PSF phase.
However, exact QMC simulations of this problem \cite{soyler2008},
in line with the results of our toy-model, predict the transition
between 2SF and PSF, at half integer filling and symmetry between
the two species, to happen at $J\approx 0.1\,|U_{ab}|$. Exploiting
the possibility of having $U_{ab}$ of the same order of $U$, this
leads to relatively large values for the critical tunneling. On
the other hand, careful analysis about the critical temperature
and entropy for the formation of the PSF phase, as recently done
in \cite{capogrosso2009}, are required.

Our model provides an oversimplified description of the system. We
believe however that it includes correctly the fundamental
ingredients of the physics involved.  A more quantitative analysis
based on exact numerical calculations, including the effects of
two-species unbalance on the formation of the paired phases, will
be subject of future work.

We acknowledge financial support by MIUR PRIN 2007 . The authors
thank F.~Becca, I.~Bloch, J.~Catani, F.~Minardi, M.~Modugno,
B.~Svistunov, and E.~Taylor, for interesting discussions.


\begin{thebibliography}{99}
\bibitem{footnote2} In the infinite system, the SF phase is defined by the
conditions $\langle a \rangle, \langle b \rangle \neq 0$ and
$\langle ab \rangle - \langle a \rangle \langle b \rangle=0$,
while the conditions for the PSF phase are $\langle a \rangle =
\langle b \rangle=0$ and $\langle ab \rangle - \langle a \rangle
\langle b \rangle \neq 0$.
\end{thebibliography}
\end{document}